\documentstyle[aps]{revtex}

\input epsf         
\epsfverbosetrue    

\begin{document}
\twocolumn[\hsize\textwidth\columnwidth\hsize\csname@twocolumnfalse%
\endcsname

\title{Transport and the Order Parameter of Superconducting Sr$_2$RuO$_4$}
\author{W. C. Wu$^1$ and R. Joynt$^2$}
\address{$^1$Department of Physics, National Taiwan Normal University,
Taipei 11650, Taiwan}
\address{$^2$Department of Physics, University of Wisconsin-Madison, Madison, WI 53706}

\date{\today}

\maketitle
\draft

\begin{abstract}
Recent experiments make it appear more likely that the order parameter of the
unconventional superconductor Sr$_2$RuO$_4$
has a spin-triplet $f$-wave symmetry.
We study ultrasonic absorption and thermal conductivity of
superconducting Sr$_2$RuO$_4$ and fit to the recent data for various
$f$-wave candidates. It is shown that
only $f_{x^2-y^2}$-wave symmetry can account qualitatively
for the transport data.
\end{abstract}

\vspace{0.1cm}
]

The high-$T_{c}$-analog compound Sr$_{2}$RuO$_{4}$ poses a
fundamental challenge to the study of unconventional superconductivity \cite%
{maeno01}. \ Clean samples become superconducting at about $T_{c}=1.5K$
and the thermodynamics show that it is unconventional. The
precise nature of its superconducting order parameter remains elusive,
however. \ The simplest spin triplet case is the
time-reversal-symmetry-breaking odd-parity $p$-wave state which has an order
parameter ${\bf d}=\hat{{\bf z}}(k_{x}+ik_{y})$ \cite{rice95}. \
Spin-triplet character is consistent with the Knight shift \cite{ishida98}
and neutron scattering \cite{duffy00} measurements which indicate that the
spin susceptibility shows no obvious suppression below $T_{c}$. Moreover, $%
\mu $SR experiment \cite{luke98} confirms that the order parameter does
indeed break time-reversal symmetry. The central feature of this $p$-wave
order parameter is that it exhibits finite gap all over the Fermi surface
(FS). \ However, more recent thermodynamic and transport measurements at low
temperatures have found powers laws in the temperature dependences. \ This
includes specific heat $C_{V}\propto T^{2}$ \cite{nishizaki00}, penetration
depth $\Delta \lambda \propto T^{2}$ \cite{bonalde00}, Ru spin-lattice
relaxation rate $1/T_{1}\propto T^{3}$ \cite{ishida00}, ultrasonic
attenuation $\alpha \propto T^{3}$ \cite{matsui01,lupien01}, and thermal
conductivity $\kappa \propto T^{2}$ \cite{tanatar01,izawa00}. \ These powers
laws in $T$ are consistent with line nodes in the order parameter. \

The combination of all experiments therefore favors the spin-triplet
odd-parity {\it f}-wave symmetries which feature both time-reversal symmetry
breaking and line nodes. \ Maki and Yang \cite{maki99} have proposed an $f$%
-wave model with a superconducting order parameter ${\bf d(}T{\bf )}=\Delta
(T)\hat{\,{\bf z\,}}k_{z}(k_{x}+ik_{y})^{2}$ (the so-called ``3D'' model)
for Sr$_{2}$RuO$_{4}$. \ This 3D gap has line nodes where the basal
plane ($k_{z}=0$) intersects the FS.  It is
similar to one of the leading candidates for the
order parameter of UPt$_{3}$ (known as the $E_{2u}$ model in that hexagonal
system \cite{graf00-2}). \ Graf and Balatsky \cite{graf00} have proposed
another candidate $f$-wave order parameter ${\bf d}=\Delta (T)\,\hat{{\bf z}}%
(k_{x}+ik_{y})k_{x}k_{y}$, which we will call the $f_{xy}$ model. \ This
order parameter has line nodes perpendicular to the basal plane at %
where the planes $k_{x}=0$ and $k_{y}=0$ intersect the FS.  This nodal
structure is analogous to the
$p$-wave model ${\bf d}=\Delta (T)\,\hat{{\bf z}}(\sin k_{x}+i\sin k_{y})$
proposed by Miyake and Narikiyo \cite{miyake99}. \ It is suggested that $%
f_{xy}$-wave order parameter could arise from a pairing interaction mediated
by antiferromagnetic spin fluctuations with peak intensities at the
incommensurate wavevectors ${\bf q}=(\pm 2/3\pi ,\pm 2/3\pi )$. The latter
has been confirmed by inelastic neutron scattering \cite{sidis99}. \ Finally
there is the $f_{x^{2}-y^{2}}$-wave order parameter ${\bf d}=\Delta (T)\,\hat{{\bf z}}%
(k_{x}+ik_{y})(k_{x}^{2}-k_{y}^{2})$ with line nodes
where the planes $k_x = \pm k_y$ intersect the FS.
This has been suggested \cite{dahm00} partly by analogy to
high $T_{c}$ cuprates where a similar
$x^{2}-y^{2}$ (but with spin-singlet) gap structure has been conclusively
established. \ It is worth noting that the above three models all account
reasonably well for the specific heat and superfluid density data \cite{dahm00,won00}.

The structure of all of these gaps in momentum space is rather complicated.
\ Since all thermodynamic and transport quantities involve Fermi surface
averages, it is not easy to see how they may be distinguished
experimentally. \ In the study of heavy-fermion superconductors, it has been
found that thermal conductivity and ultrasonic attenuation can be key tools
in this situation. \ These quantities can be studied as a function of
propagation direction and, in the case of the attenuation, also of
polarization dependence. \ The key point is that this directional
sensitivity can not only test for the presence of point or line nodes, but
can also determine the positions of nodes and the orientations of nodal
lines. \ These experiments \cite{coffey85} offer the best chance to
distinguish between order parameters that have different nodal structures
but share similar thermodynamic properties.

In this paper, we calculate the thermal conductivity and the transverse and
longitudinal ultrasonic attenuation of Sr$_{2}$RuO$_{4}$. \ We consider the
three major candidates of $f$-wave symmetry mentioned above, namely the 3D, $%
f_{xy}$, and $f_{x^{2}-y^{2}}$ models. It will be shown that only the $%
f_{x^{2}-y^{2}}$ model can account for the data even qualitatively. \ The
in-plane $k_{x}^{2}-k_{y}^{2}$ symmetry offers an intriguing similarity to
the high-T$_{c}$ order parameter. \ If this gap structure is indeed manifested
in Sr$_{2}$RuO$_{4}$ it would suggest the presence of
anti-ferromagnetic (AF) fluctuations in the ruthenate superconductors,
perhaps peaked at or near ${\bf q}=(\pi ,\pi )$ (such as one finds in the
cuprates). \ The transport data thus suggests that AF fluctuation may also
exist in ruthenates.

Ultrasonic attenuation experiments in Sr$_{2}$RuO$_{4}$ have been done in
the hydrodynamic regime where the sound frequency $\omega $ satisfies $%
\omega <<1/\tau $, where $\tau $ is the electronic relaxation time \cite%
{lupien01}. \ In this case the Boltzmann equation is appropriate, and the
attenuation $\alpha_{ij} $ for sound propagation along the direction
$\hat{q}\parallel \hat{i}$
with polarization $\hat{\varepsilon}\parallel \hat{j}$ is \cite{moreno96}:

\begin{eqnarray}
\alpha _{ij}(T)&\propto& \int_{0}^{\infty}
d\omega \left[ -{\frac{\partial F(\omega )}{\partial \omega }}\right] \tau
_{s}(\omega ,T)\nonumber\\
&\times&\left\langle {\rm Re}\left( {\frac{\sqrt{\omega ^{2}-
|{\bf d}_{{\bf k}}|^{2}}}{\omega }}[(\hat{i}\cdot \hat{{\bf k}})
(\hat{j}\cdot \hat{{\bf k}})-{1\over d}\delta_{ij}]^{2}\right) \right\rangle.
\label{eq:alpha}
\end{eqnarray}
Here $d$ denotes the dimensionality,
$F$ is the Fermi function, and the angle brackets indicate an FS average.
Note the second term inside the square bracket vanishes for the transverse case.
Similarly the thermal conductivity in the $i$th
direction when a temperature gradient is imposed along the $i$th direction
is computed using:

\begin{eqnarray}
\kappa_{i}(T)&\propto& {\frac{1}{T}}\int_0^\infty d\omega ~\omega^2\left[-{%
\frac{\partial F(\omega)}{\partial \omega}}\right] \tau_s(\omega,T)
\nonumber \\
&\times&\left\langle{\rm Re} \left({\frac{\sqrt{\omega^2-|{\bf d}_{{\bf k}%
}|^2}}{\omega}} (\hat{\imath}\cdot\hat{{\bf k}})^2 \right)\right\rangle.
\label{eq:kappa}
\end{eqnarray}

The FS of Sr$_{2}$RuO$_{4}$ has been determined by de Haas-van Alphen
experiments. \ It has three sheets denoted by $\alpha $, $\beta $, and $%
\gamma $ \cite{mackenzie96}. \ These are in good agreement with LDA
calculations \cite{bands}. \ The bands are two-dimensional, a feature
similar to cuprates. Among these three bands, the $\gamma $ band has the
largest FS and has the highest density of states (DOS) at the Fermi energy.
\ For simplicity, we use a single cylindrical FS for $f_{xy}$ and $%
f_{x^{2}-y^{2}}$ models and a three-dimensional spherical FS for the 3D
model. \ We shall comment on the effects of FS anisotropy below. \ We will
assume that the momentum dependence of the gap is independent of
temperature, which leaves only the overall magnitude $\Delta (T).$ \ The
temperature dependence of the gap in Sr$_{2}$RuO$_{4}$ has not been directly
measured. However, one may deduce the power law at low temperatures from the
pattern of nodes and near $T_{c\text{ }}$the power is given by mean field
theory. For line nodes, a reasonable interpolation for the temperature
dependence is $\Delta (T)=\Delta (0)[1-(T/T_{c})^{3}]^{1/2}$. \ This form
has only one parameter $\Delta (0)$, reducing the arbitrariness in the
fitting procedure.

Inspection of Eqs.~(\ref{eq:alpha}) and (\ref{eq:kappa}) shows that the only
remaining unknown is the relaxation time $\tau _{s}(T)$. \ Our
procedure will be to use the
thermal conductivity to determine the relaxation time and the
magnitude of the ground state gap $\Delta (0)$, then use these to
constrain the fits of the attenuation.

\begin{figure}
\mbox{
\epsfxsize=0.9\hsize
{\epsfbox{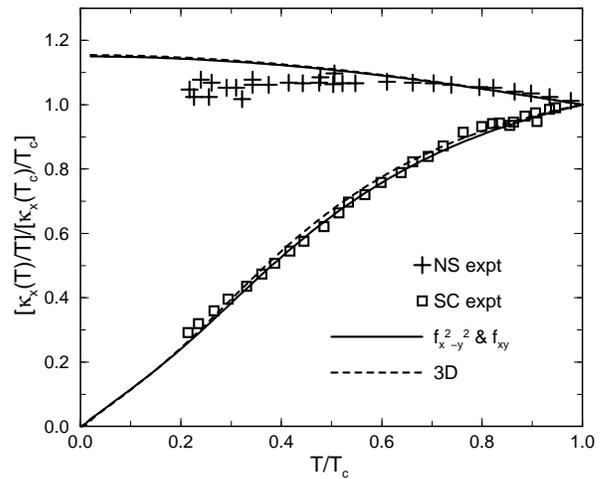}}}
\vspace{0.2cm}
\caption{Theoretical fits of thermal conductivity against the data of %
\protect\cite{tanatar01} in both the normal (upper) and superconducting
(lower) states. The fits are equally well for the three candidates. The
fitting parameters are given in the text.}
\label{fig1}
\end{figure}

In Fig.~\ref{fig1}, we first show the best fits against the thermal
conductivity $\kappa_x$ along the [100] direction in
the normal and superconducting states.
\ The normal state is achieved experimentally by applying a sufficiently
strong external magnetic field. \ The data are those of Tanatar {\em et al.} %
\cite{tanatar01}. \ In the normal state at $0.5T_c\alt T \leq T_c$,
the scattering rate can be well
described by an empirical form: $1/\tau _{n}=\Gamma _{0}(1+aT^{2}/T_{c}^{2})$
with $\Gamma _{0}\simeq 0.01\pi T_{c}$ and $a=0.15$. \ That is, the inelastic
scattering rate is observed to be roughly $15\%$ of the elastic part at $%
T_{c}$. \ For the effective scattering rate in the superconducting state, we
take the same temperature dependence as in the normal state. \ This is
clearly appropriate for $T\alt T_{c}$. \ At lower temperatures $T\ll T_c$, due to
the opening of the gap, the inelastic rate could decrease faster than $T^{2}$.
However, since the elastic scattering rate dominates over the inelastic
part, the exact temperature dependence of inelastic rate is not so
important. \ (This is one big difference between this system and the heavy
fermion and high-$T_c$ materials.) \ Fitting to the superconducting thermal conductivity
data then fixes the other parameter $\Delta (0)$. It turns out that all
three theoretical candidates can fit the data for $\kappa_x$
in the superconducting state.
\ There is no noticeable difference in goodness-of-fit for the different
models. \ The parameters determined by the fits are $\Delta (0)=1.65T_{c}$
for $f_{x^{2}-y^{2}}$ model, $\Delta (0)=3.3T_{c}$ for $f_{xy}$ model, and $%
\Delta (0)=4.05T_{c}$ for 3D model. \ The difference in size between these
various values is not significant, since each gap form has a different relation
between $\Delta (0)$ and the root-mean-square FS average gap. \ It is the
latter quantity that is more important in determining the overall scattering
strength.

\begin{figure}
\mbox{
\epsfxsize=0.9\hsize
{\epsfbox{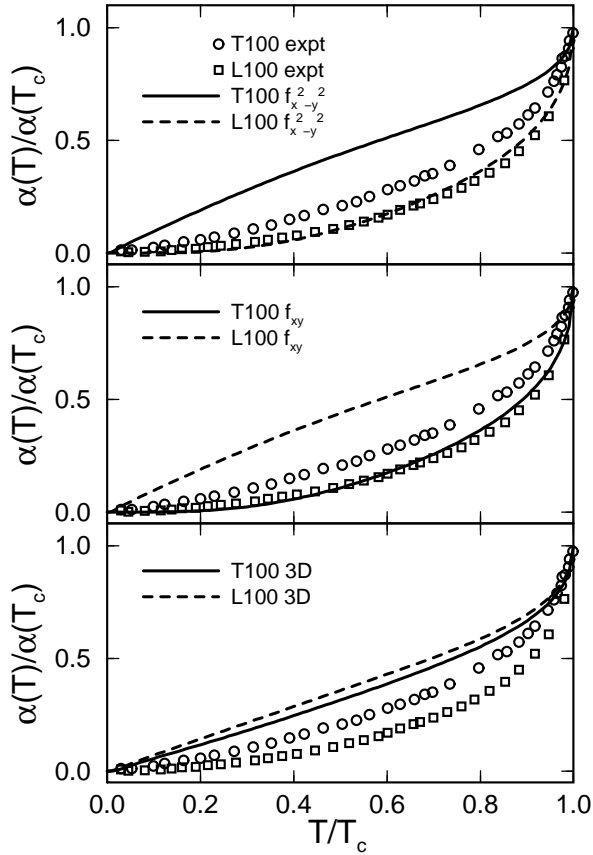}}}
\vspace{0.2cm}
\caption{Theoretical calculations of ultrasonic attenuation for $f_{x^2-y^2}$%
, $f_{xy}$, and 3D models (from top to bottom). The parameters used are
identical to those in Fig.~\ref{fig1}. Experimental data is from Ref.~%
\protect\cite{lupien01}.}
\label{fig2}
\end{figure}

The parameters determined by the fit in Fig.~\ref{fig1} are now used to
calculate the transverse and longitudinal ultrasonic attenuation {\it %
without any adjustable parameters} (Fig.~\ref{fig2}) for all three models.

As seen in Fig.~\ref{fig2}, the experiment exhibits the very clear feature
that the transverse T100 $\alpha $ is systematically higher than the
longitudinal L100 $\alpha $ when scaled to the normal state values at $T_{c}$%
. \ Note that the polarization vector for the transverse sound is also in
the $x-y$ plane. \ This sets a crucial constraint on theoretical models. \
\

Of the three candidate gaps, only the $f_{x^{2}-y^{2}}$-wave has this
property. \ The reason for this can be seen semi-quantitatively as follows. \
Inspection of Eq.~(\ref{eq:alpha}) shows that one big difference between the
transverse and longitudinal cases is a factor of
$\hat{k}_{x}^{2}\hat{k}_{y}^{2}$, and $(\hat{k}_{x}^2-1/d)^2$, respectively.
The other factors restrict the integral to the FS, and, as a function of temperature,
weight the points on the FS according to how small the gap is. \ At the
lowest temperatures, only the nodal regions contribute. \ Thus we may think
of the integral as being over an effective FS located near the nodes, with a
polarization-dependent weighting factor. \ The 3D gap has an effective FS
that is a circle at the equator $k_{z}=0.$ \ For this geometrical object, we have
$\langle \hat{k}_{x}^{2} \hat{k}_{y}^{2}\rangle=\langle
(\hat{k}_{x}^2-1/d)^2\rangle \equiv 1/8$ with $d=2$.
The angle brackets indicate an average over
the effective FS. \ Thus the average for longitudinal polarizations is
about the same as that for transverse. \ For the $f_{xy}$ state, the  effective FS
is a set of two lines at ${k}_{x}=0$ and ${k}_{y}=0$, and we
have\ $\langle \hat{k}_{x}^{2} \hat{k}_{y}^{2}\rangle \equiv 0$, while
$\langle (\hat{k}_{x}^{2}-1/d)^2\rangle $ is finite ($d=2$), favoring longitudinal
polarization. \ In contrast, for the $f_{x^{2}-y^{2}}$ state, the effective
FS\ is at ${k}_{x}=\pm {k}_{y}$. \ The averages
therefore lead to $\langle \hat{k}_{x}^{2}\hat{k}_{y}^{2}\rangle$
finite, while $\langle (\hat{k}_{x}^2-1/d)^2\rangle \equiv 0$ ($d=2$ again),
favoring transverse attenuation.
These averages show that the $f_{x^{2}-y^{2}}$ will naturally have more
transverse attenuation than its competitors.  Although we have plotted only
one set of curves for each of the states, this is a general feature.
The fact that the attenuation data favor $f_{x^{2}-y^{2}}$ over other
candidates was already noted by
Lupien {\em et al.} \cite{lupien01} without calculation.

Since the data in Figs.~\ref{fig1} and \ref{fig2} are taken on different
samples ($T_c=1.44$K and 1.37K, respectively),
one does not expect a quantitatively good fit in Fig.~\ref{fig2}
using the same parameters in Fig.~\ref{fig1}.
At first glance, our results of $\alpha$ for $f_{x^2-y^2}$ and
$f_{xy}$ waves are qualitatively similar to
Graf and Balatsky's \cite{graf00} when
a resonant scattering limit ($\delta=90^{\rm o}$) is taken
under the self-consistent $T$-matrix approximation.
Lupien {\em et al.} \cite{lupien01} pointed out
that while the sound attenuation data favors $f_{x^2-y^2}$ wave,
the low $T$ power law is much lower than the calculated
result of Graf and Balatsky for the longitudinal case.
In contrast, as shown in Fig.~\ref{fig2}, our
calculation based on a phenomenological input of scattering time
gives satisfactory good fit for the longitudinal case including
the low $T$ power law. It also leads to better fit for
the transverse case especially at $T\alt T_c$,
compared to Graf and Balatsky's almost linear in $T$ result.
The still rather large discrepancy between
the theoretical fits and the experimental data (especially for the T100 mode)
could be due to the order parameter which is more involved
than having the pure $f_{x^2-y^2}$ symmetry.
This remains to be studied in more details.


It is important to note one major difference between
our phenomenological treatment and the self-consistent
$T$-matrix approximation used by Graf and Balatsky.
In the unitary limit, the usual
$T$-matrix approximation for the impurity scattering
will assume a strongly frequency dependent
scattering rate $1/\tau_{\rm imp}\sim 1/\omega$ in the
superconducting state \cite{graf00}.
In contrast, we have adopted a constant elastic scattering rate
({\em i.e.}, $\Gamma_0$).
Our phenomenological input relies truly on the good fit to the
thermal conductivity data in the superconducting state (see Fig.~\ref{fig1}).

In-plane thermal conductivity measured by Izawa {\em et al.}
\cite{izawa00} with a rotating in-plane magnetic field has shown a four-fold
pattern with a minimum when ${\bf H}\parallel \lbrack 110]$. This angle
dependence is consistent with the  $f_{x^{2}-y^{2}}$ and 3D models but rules
out the $f_{xy}$ model \cite{izawa00}. \ In terms of variation of the
amplitude, Izawa {\em et al. }claimed that the state ${\bf d}=\Delta (T)\,%
\hat{{\bf z}}(k_{x}+ik_{y})(\cos ck_{z}+|b|)$, originally proposed
by Hasegawa {\em et al.} \cite{hasegawa00}, is more consistent with the
data based on the approach in Refs.~\cite{won00,dahm00}. \ Here $c$ is the
spacing between two adjacent Ru layers and $|b|\leq 1$. \ This model has a
circular-like line node parallel to the basal plane with any $k_{z}$
satisfying $-1\leq k_{z}c\leq 1$. This gap is symmetrical for any given
parallel ${\bf k}$ plane and will lead to a sound attenuation very similar
to that of the 3D model above. \ Thus it also does not produce the proper
polarization dependence.

Finally we comment on the effect of Fermi surface anisotropy. \ While in Sr$%
_{2}$RuO$_{4}$ the FS has multiple sheets and there seems to be a coupling
between the $\alpha $ and $\beta $ bands, the analysis of transport data
given here will not be significantly modified providing that the $D_{4h}$
symmetry of the crystal is preserved. \ Moreover, our result for the 3D model
stands if the spherical FS is replaced by an ellipsoidal one that has
circular projection parallel to the basal plane. \ As long as transport data
is concerned, it is likely that interband effects are unimportant.

In conclusion, the transport data of Sr$_{2}$RuO$_{4}$ presently give strong
preference to the $f_{x^{2}-y^{2}}$-wave order parameters with perpendicular
line nodes at $k_{x}=\pm k_{y}$. \ A sharper comparison of theory and
experiment will be possible if the data for $\alpha $ and $\kappa $ is taken
on the same sample. Furthermore, we suggest that the out-of-plane
transverse sound attenuation
measurement should be performed under the polarization
$\hat{q}\parallel \hat{x}$ and
$\hat{\varepsilon}\parallel \hat{z}$. This will shed more light
on the information of the location of line nodes and the
overall nodal structure.

Financial support from NSC of Taiwan
(Grant No. 89-2112-M-003-027) and the NSF under
the Materials Theory program, Grant No. DMR-0081039
is acknowledged.


\begin{references}

\bibitem{maeno01}
{Y. Maeno, T.M. Rice, and M. Sigrist}, Physics Today {\bf 54 (1)},  42  (2001).

\bibitem{rice95}
{T.M. Rice and M. Sigrist}, J. Phys.: Condens. Matter {\bf 7},  L643  (1995).

\bibitem{ishida98}
{K. Ishida, H. Mukuda, Y. Kitaoka, K. Asayama, Z.Q. Mao, Y. Mori, and Y.
  Maeno}, Nature {\bf 396},  658  (1998).

\bibitem{duffy00}
{J.A. Duffy, S.M. Hayden, Y. Maeno, Z.Q. Mao, J. Kulda, and G.J. McIntyre},
  Phys. Rev. Lett. {\bf 85},  5412  (2000).

\bibitem{luke98}
{G. M. Luke, Y. Fudamoto, K.M. Kojima, M.I. Larkin, J. Merrin, B. Nachumi, Y.J.
  Uemura, Y. Maeno, Z.Q. Mao, Y. Mori, H. Nakamura, and M. Sigrist}, Nature
  {\bf 394},  558  (1998).

\bibitem{nishizaki00}
{S. Nishizaki, Y. Maeno, and Z.Q. Mao}, J. Phys. Soc. Jpn. {\bf 69},  572
  (2000).

\bibitem{bonalde00}
{I. Bonalde, B.D. Yanoff, M.B. Salamon, D.J. Van Harlingen, E.M.E. Chia, Z.Q.
  Mao, and Y. Maeno}, Phys. Rev. Lett. {\bf 85},  4775  (2000).

\bibitem{ishida00}
{K. Ishida, H. Mukuda, Y. Kitaoka, Z.Q. Mao, Y. Mori, and Y. Maeno}, Phys. Rev.
  Lett. {\bf 84},  5387  (2000).

\bibitem{matsui01}
{H. Matsui, Y. Yoshida, A. Mukai, R. Settai, Y. Onuki, H. Takei, N. Kimura, H.
  Aoki, and N. Toyota}, Phys. Rev. B {\bf 63},  R60505  (2001).

\bibitem{lupien01}
{C. Lupien, W.A. MacFarlane, C. Proust, L. Taillefer, Z.Q. Mao, and Y. Maeno,
  Phys. Rev. Lett. {\bf 86}, June 11, 2001 issue.}

\bibitem{tanatar01}
{M. A. Tanatar, S. Nagai, Z.Q. Mao, Y. Maeno, and T. Ishiguro}, Phys. Rev. B
  {\bf 63},  64505  (2001).

\bibitem{izawa00}
{K. Izawa, H. Takahashi, H. Yamaguchi, Y. Matsuda, M. Suzuki, T. Sasaki, T.
  Fukase, Y. Yoshida, R. Settai, and Y. Onuki}, Phys. Rev. Lett. {\bf 86},
  2653  (2001).

\bibitem{maki99}
{K. Maki and G. Yang}, Fizika {\bf 8},  345  (1999).

\bibitem{graf00-2}
{M. J. Graf, S.-K. Yip, and J. A. Sauls}, Phys. Rev. B {\bf 62},  14393
  (2000).

\bibitem{graf00}
{M.J. Graf and A.V. Balatsky}, Phys. Rev. B {\bf 62},  9697  (2000).

\bibitem{miyake99}
{K. Miyake and O. Narikiyo}, Phys. Rev. Lett. {\bf 83},  1423  (1999).

\bibitem{sidis99}
{Y. Sidis, M. Braden, P. Bourges, B. Hennion, S. Nishizaki, Y. Maeno, and Y.
  Mori}, Phys. Rev. Lett. {\bf 83},  3320  (1999).

\bibitem{dahm00}
{T. Dahm, H. Won, and K. Maki, cond-mat/0006301.}

\bibitem{won00}
{H. Won and K. Maki}, Europhys. Lett. {\bf 52},  427  (2000).

\bibitem{coffey85}
{L. Coffey, T.M. Rice, and K. Ueda}, J. Phys. C {\bf 18},  L813  (1985).

\bibitem{moreno96}
{J. Moreno and P. Coleman}, Phys. Rev. B {\bf 53},  R2995  (1996).

\bibitem{mackenzie96}
{A. P. Mackenzie, S.R. Julian, A.J. Diver, G.J. McMullan, M.P. Ray, G.G.
  Lonzarich, Y. Maeno, S. Nishizaki, and T. Fujita}, Phys. Rev. Lett. {\bf 76},
   3786  (1996).

\bibitem{bands}
{T. Oguchi, Phys. Rev. B {\bf 51}, 1385 (1995); D. J. Singh, Phys. Rev. B {\bf
  52}, 1358 (1995).}

\bibitem{hasegawa00}
{Y. Hasegawa, K. Machida, and M. Ozaki}, J. Phys. Soc. Jpn. {\bf 69},  336
  (2000).

\end{references}

\end{document}